\documentstyle[twocolumn,aps,prl,epsf]{revtex}
\input epsf

\begin{document}
\draft
\title{Mass Extinction in a Simple Mathematical Biological Model}
\author{Kei Tokita\cite{byline}}
\address{Department of Chemistry and Chemical Biology,
         Harvard University, 12 Oxford Street, Cambridge MA 02138}
\author{Ayumu Yasutomi}
\address{Suntory and Toyota International Centres
   for Economics and Related Disciplines\\
   London School of Economics and Political Science,
   London WC2A 2AE, UK}
\date{\today}
\maketitle
\begin{abstract}
Introducing the effect of extinction into the so-called replicator
equations in mathematical biology, we construct a general model of
ecosystems.  The present model shows mass extinction by its own {\it
extinction dynamics} when the system initially has a large number of
species ({\it diversity}). The extinction dynamics shows several
significant features such as a power law in basin size distribution,
induction time, etc. The present theory can be a mathematical
foundation of the {\it species-area effect} in the paleontologic
theory for mass extinction.
\end{abstract}

Mechanisms of mass extinction of species in ecosystems have been
studied by a number of
researchers\cite{Raup_1991,Maynard_Smith_1989}.  Their conclusions
can be divided into two categories, one emphasizing exogenous
shocks\cite{Hedges_etal,Alvarez_et_al_1980,Hallam_1990,%
Coffin_Eldholm_1993} and the other, endogenous
causes\cite{Van_Valen_1973,Bak_Sneppen_1993,%
Newman_Roberts_1995}. Building on both views, we construct a
general mathematical biological model of ecosystems. This model reflects the
former view, e.g., the situation where several biotas which have
been separated from each other for a long time are suddenly
integrated into a larger ecological network (biotic fusion) by
some exogenous shock\cite{Vermeij_1991,Gilpin_1994}.
We assume that the interaction coefficients for this newly
produced ecosystem can be written in the form of a random
matrix\cite{Gardner_Ashby_1970,May_1972,Roberts_1974,%
Opper_Diederich_1992}. Also, following the latter view, we adopt
the concept of an `extinction threshold', which we introduce into
the replicator equations\cite{Hofbauer_Sigmund_1988} of population
dynamics. Using this replicator equation model with random
interaction and the extinction threshold, we find several
significant new features characterizing the nature of mass
extinction.

We investigate the following $N_I$ dimensional ordinary
differential equations:
\begin{equation}
\frac{{\rm d}x_i(t)}{{\rm d}t} = x_i(t)
      \left(\sum_{j=1}^{N_I} a_{ij}x_j(t)
     -
      \sum_{j,k=1}^{N_I} a_{jk}x_j(t)x_k(t)\right)\label{replicator_eq}
\end{equation}
\begin{equation}
\sum_{i=1}^{N_I} x_i(t) = 1\qquad\qquad (0\le x_i(t) \le 1).
                               \label{constraint}
\end{equation}
These equations, the {\it replicator equations}, are generally used to
describe the evolution of self-replicating entities, so-called
replicators\cite{Dawkins_1982}. The equations are also termed the {\it
game dynamical equations} in the game theory. Moreover, they are
equivalent to the $N_I-1$ dimensional Lotka-Volterra
equations\cite{Hofbauer_Sigmund_1988}. Therefore, we will use the term
{\it species} when referring to replicators hereafter.  The variable
$x_i$ denotes the {\it population density} of the species $i$.  $N_I$
denotes the initial number of species, that is, the initial value of
the diversity.
The $(i,j)$-th element of the matrix $A=(a_{ij})$ determines the
effect of species $j$ on the growth rate of species $i$. Here we
use $a_{ii}=-1$ for intraspecies interaction coefficients and
assign the interspecies ones $a_{ij}\, (i\neq j)$ as
time-independent Gaussian random numbers with mean $0$ and
variance $v$. This assumption of random interactions is based on
the hypothesis that a biotic fusion reorganizes species
relationships in a random fashion\cite{Gilpin_1994}. This kind of
ecosystem with random interaction also can be produced by the
so-called {\it species-area effect} that paleontologists have
asserted to be a trigger of mass extinction\cite{Hallam_1994}. For
example, because the species-area effect may be caused by
declining sea levels, which confines many biologically-isolated
species to a narrow area, drives them into competition and,
eventually, brings biotic fusion.  We also believe that the random
interaction model is important as a first step of understanding
the behavior of a large ecosystem with many species interacting in
a complex way. For such a random interaction model, the local
stability condition has been
clarified\cite{Gardner_Ashby_1970,May_1972,Roberts_1974} in the
$N_I\to\infty$ limit. However, the global behavior is hardly
treated analytically, because the equations are highly nonlinear
and the dynamics often shows complex behavior, such as {\it
heteroclinic orbits} \cite{May_et_al_75,Chawanya_95} or {\it
chaos}\cite{Gilpin_1979,Arneodo_et_al_1980}, even at a small
degree of freedom ($N_I \ge 4$).

Here we should note that extinction is not well-defined in the above
model with a large $N_I$ because of the heteroclinic orbits. 
The reason is that even though a heteroclinic orbit approaches a
{\it saddle} where some species are extinct, the population
densities never reach exactly zero and the orbit eventually leaves
for another saddle where the population densities
revive. Moreover, this transition among saddles continues
cyclically or chaotically.
In this sense, heteroclinic orbits have never been believed to be
biologically siginificant.

Considering the above problem, we introduce a parameter $\delta$,
the extinction threshold, to the dynamics
(\ref{replicator_eq})-(\ref{constraint}) : at each discrete time
step, the population density $x_k$ is set to zero if this quantity
becomes less than $\delta$. The population densities of the
surviving species \{$x_i$\}$\, (i\neq k)$ are then renormalized to
satisfy $\sum_{i\neq k}x_i=1$. This renormalization implies that
the niche of an extinct species is divided among the
survivors. The diversity decreases through the above process, and
we denote its value by $N$. Thus, the present model can be
interpreted as a dynamical system whose degree of freedom is a
time-dependent variable.  Although this time-dependent degree of
freedom is inevitable not only in population
dynamics\cite{Tregonning_Roberts_1979,Ginzburg_et_al_1988} but
also in many fields, such a highly non-linear model has never been
systematically analyzed.

\begin{figure}[t]
  \epsfxsize=7.75cm
  \epsfbox{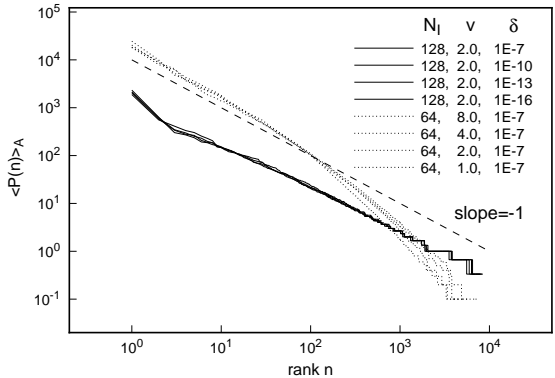}
   \caption{Basin-size distributions for (a)
  $N_I=64$ sampled from 100000 initial states and averaged over $10$
  samples of $A$, (b) $N_I=128$ from 20000 initial states and $3$
  samples of $A$.}
 \label{fig1}
\end{figure}

Whenever given a set of parameters $A$, $\delta$ and the initial
diversity $N_I\equiv N(t=0)$, the initial state $\{x_i(0)\}$
evolves until a steady state is achieved. In this state extinction
no longer occurs, and there remains a stable subecosystem with a
comparatively small number of surviving species ({\it core
species}) $N_F\; (\leq N_I)$. In this state, although almost all
orbits converge to an equilibrium point, we also find periodic
orbits. Chaotic orbits are very rare. Heteroclinic orbits are
never achieved because the existence of the finite $\delta$
prohibits any orbit from approaching a saddle.

This introduction of $\delta$ also introduces a finite size effect
into the replicator equations, because $\delta$ coincides with a
minimum unit of reproduction of each species, and its reciprocal
$1/\delta$ corresponds to the permissible population size of the
ecosystem. Let us refer to this kind of dynamics as {\it
extinction dynamics} (ED). By the series of extensive numerical
simulations, we investigate novel features of ED, especially the
dependence of ED on three parameters: $N_I$, $v$ and $\delta$.

From the view point on the random system theory, it is important
to observe a typical behavior of ED by executing {\it random
average} of quantities over samples of a random matrix
$A$. Hereafter, we will in general write this average as $\langle ... 
\rangle_A$.

The first result of this Letter concerns basin-size distribution
of ED which has a large number of basins of attraction. Here, we
identify each `attractor' only by composition of core species, not
by its trajectory. In other words, even if there coexist more than
one isolated attractor in a system of core species, we do not
discriminate these attractors and we regard them to be in one
basin of `attraction'. The reason is that in ED such coexistence
is rare and this classification of basins of attraction also
agrees with a classification of subecosystems appearing by ED.

In order to obtain the basin-size distribution, we (a) iterate ED
starting from a sufficient number of random initial states in a
system with same parameters and a same random matrix $A$, (b)
count basin size as the number of initial states which converge to
each `attractor', and (c) make a rank-size distribution $S(n)$,
where the natural number $n$ denotes the rank of each basin and
can reach the total number of `attractors' found in the
simulation. Moreover, the above process is iterated for a
sufficient number of random matrices $A$ with same $v$ and we
finally obtain a basin-size distribution $\langle S(n)\rangle_A$
for a parameter set. $\langle S(n)\rangle_A$'s for various
parameter sets are shown in Fig.~1.

We can see the significant characteristic that $\langle
S(n)\rangle_A$ follows the power law. This indicates that the
phase space of ED is divided in such a way that the size of each
basin of attraction resembles each term of a geometric
series. Moreover, each exponent of the power depends only on
$N_I$, not on $\delta$ nor $v$. The independence of $\delta$
strongly suggests that the basin-size distribution of the original
replicator equations (ED in the limit $\delta\to 0$) also follows
the power law. This conjecture is relevant to the hierarchal
coexistence of infinitely many attractors in the replicator
equations\cite{Chawanya_96}.
The power law of rank-size
relationship with exponent near unity is often referred to as Zipf's
law\cite{Zipf_law}, known in linguistics and diverse fields.
 
\begin{figure}[tb]
  \epsfxsize=7.75cm
  \epsfbox{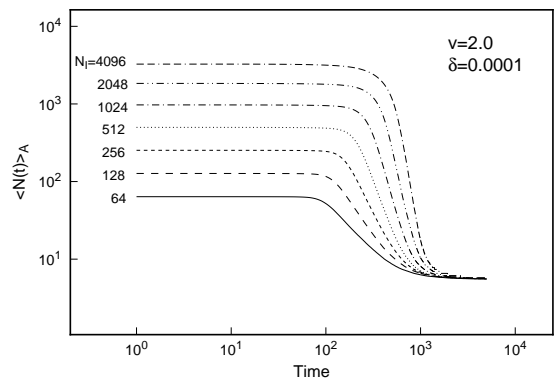}
	\caption{Extinction curves for various values of $N_I$ with
  $\delta=0.0001$ and $v=2.0$. Each curve represents an average taken
  over $1000$ samples of $A$.}  
 \label{fig2}
\end{figure}

Figure 2 shows the second result of this Letter: average diversity
as a function of time, $\langle N(t) \rangle_A$ ({\it the
extinction curve}). Two significant characteristics can be seen
from this figure. First, the average final diversity $\langle
N_F\rangle_A$ is independent of $N_I$. This result implies that no
matter how large the diversity of initial species, the average
diversity of species in the final state is small in comparison.
That is $N_F\ll N_I$. In other words, when a large random
ecosystem emerges as a result of biotic fusion, a mass extinction
of `size' $N_I-\langle N_F\rangle_A$ will occurs. Secondly, the
avalanche of mass extinction begins after some {\it induction
time}\cite{Saito_et_al_1970} $t_I$, and ends in each case at
nearly the same time $t_R\sim 10^3(\geq t_I)$. As $N_I$ becomes
larger, $t_I$ also becomes larger and approaches $t_R$.
Therefore, for sufficiently large $N_I$, the extinction curve
shows a sharp drop at $t_I$. Such an abrupt mass extinction
occurring on a short time scale is highly relevant to the notion
of {\it punctuated equilibria}\cite{Gould_Eldredge_1993}.

\begin{figure}[tb]
  \epsfxsize=7.75cm
  \epsfbox{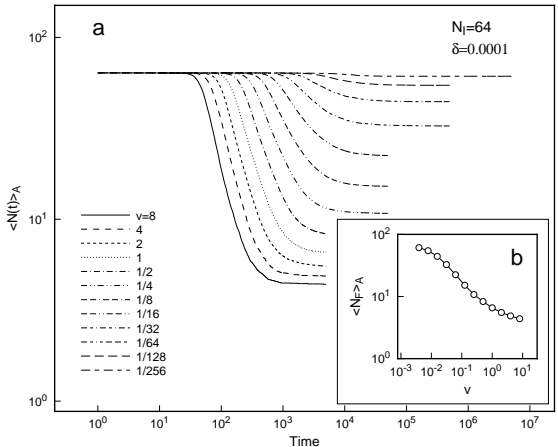}
 \caption{(a) Extinction curves for various values of $v$ with
  $N_I=64$ and $\delta=0.0001$, averaged over $1000$ samples of
  $A$. (b) The average diversity of core species $\langle N_F
  \rangle_A$ as a function of $v$. In the limit $v\to 0$, $\langle
  N_F \rangle_A \to N_I$, because this is the limit of no
  interspecies interaction($a_{ij}=0$), and in this limit the
  right hand side of Equation (\protect\ref{replicator_eq})
  becomes $0$.}
 \label{fig3}
\end{figure}

\begin{figure}[tb]
  \epsfxsize=7.75cm
  \epsfbox{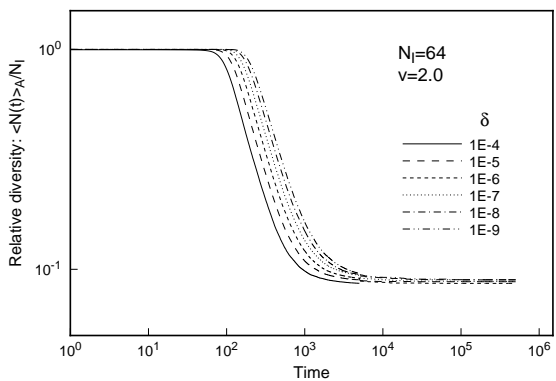}
 \caption{Extinction curves for various values of $\delta$ with
$N_I=64$ and $v=2.0$, averaged over $1000$ samples of $A$.}
 \label{fig4}
\end{figure}

Figure 3 concerns variation of extinction curves with $v$. As $v$
becomes larger, the induction time $t_I$ becomes shorter
(Fig.~3a), and $\langle N_F\rangle_A$ becomes smaller
(Fig.~3b). Consequently, when the order of the interspecies
interaction coefficients becomes large compared with the absolute
value of the intraspecies ones ($\{a_{ii}=-1\}$), the avalanche of
mass extinction begins earlier, and a smaller diversity of species
survives.
Extinction curves for several values of $\delta$ are also shown in
Fig.~4. From figures 2, 3 and 4, we can conclude that $\langle
N_F\rangle_A$ depends only on $v$, but not on $N_I$ nor $\delta$,
which contrasts the parameter dependence of $\langle
S(n)\rangle_A$ only on $N_I$.

\begin{figure}[tb]
  \epsfxsize=7.75cm
  \epsfbox{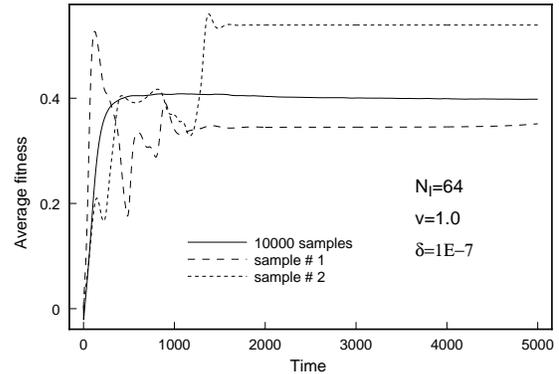}
   \caption{Time development of $\langle\bar{f}(t)\rangle_A$ over
$1000$ samples of $A$ with $N_I=64$, $v=2.0$ and $\delta=10^{-7}$. 
Two samples of average fitness $\bar{f}(t)$ are also depicted as
dotted curves}
 \label{fig5}
\end{figure}

Here we also mention the time development of the {\it average
fitness}
$\bar{f}(t)\equiv\sum_{i=1}^N\sum_{j=1}^Na_{ij}x_i(t)x_j(t)$ and
its random average $\langle\bar{f}(t)\rangle_A$.  They are
depicted in Fig.~5. It should be noted that the average fitness
takes on positive values, except during the short period in the
beginning.  The final value of $\bar{f}\sim 0.4$ is higher than
the standard deviation $\sigma\sim 0.16$ of average fitness which
is expected for a randomly generated ecosystem with same diversity
($N_F\sim 8$). Thus, more stable ecosystems are self-organized by
ED. We also observe that $\langle\bar{f}(t)\rangle_A$ does not
show monotonic increase and reaches maximum value at the time near
$t_I$.  It suggests that, in general, the average fitness shoots
up by the avalanche of extinction of low fitness species around
the induction time and settles down to the final value via
competition among core species.

\begin{figure}[tb]
  \epsfxsize=7.75cm
  \epsfbox{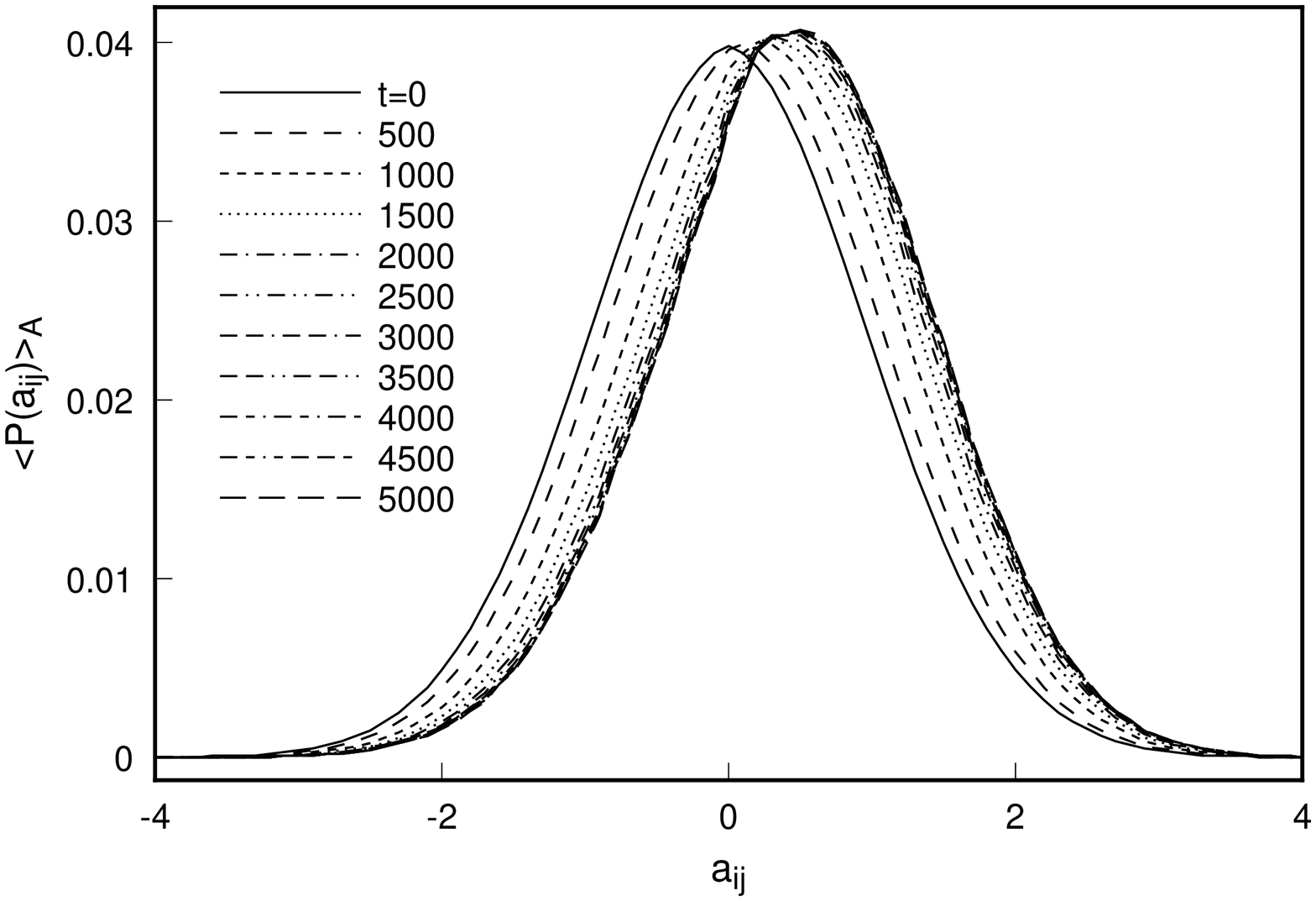}
   \caption{Snap shots of interaction coefficients distribution
averaged over $2000$ samples of $A$ with $N_I=64$, $v=2.0$ and
$\delta=10^{-7}$.}
 \label{fig6}
\end{figure}

Time development of distribution of interspecies interaction
coefficients $a_{ij} (i\neq j)$ among surviving species is
depicted in Fig.~6.  The average of $a_{ij}$ shifts to positive
value along the time, which means that the relationship among the
species becomes cooperative by ED. This also contributes the
increase of average fitness. It should be noted that the
distribution continuously holds its shape of gauss
distribution. Therefore, the interspecies interaction coefficients
of core species are still random, and various types of
relationship among core species are realized by ED.

In this Letter, we ignore any effects by immigrants or invaders, which
increase diversity, and we focus on global biotic fusion where no
species ever comes from outside. Moreover, we do not consider any
mutants because avalanche of mass extinction occurs so quickly that
any evolution of mutants never follows. By neglecting these effects,
the nature of extinction on rather short time scale is exclusively
clarified. However, by introducing the effect of the increasing
diversity, we can study the nature of ED on much larger time
scale. For example, it must be interesting problem whether ED shows
the self-organized criticality\cite{Bak_Sneppen_1993,Bak_et_al_1989}.

The present theory suggests that a biotic fusion by some external
shock will cause a mass extinction if the fusion occurs in large scale
and the interspecies interactions become random. It also can be a
mathematical biological foundation of the species-area effect because
it possibly plays a similar role like the biotic fusion. Furthermore,
our results clarify the importance of the extinction threshold,
i.e., the finite size effect for the replicator equations. 

Finally, we strongly believe in the universality of our model,
consisting of replicator equations and the finite size effect, because
the former are accepted models in many scientific fields, such as
sociobiology, prebiotic evolution of macromolecules, mathematical
ecology, population genetics, game theory and even economics, and the
existence of the latter is inevitable for such fields.

The authors would like to thank Y.~Aizawa, K.~Kaneko, T.~Ikegami
and T.~Chawanya for fruitful discussions, and G.~Sheen for her
careful reading of the manuscript. The present work is partially
supported by a Grant-in-Aid from the Ministry of Education,
Science and Culture of Japan and that from Suntory
foundation. Most of the numerical calculations were carried out on
a Fujitsu VPP500/40 in the Supercomputer Center, Institute for
Solid State Physics, University of Tokyo.

\end{document}